# High pressure synthesis of a hexagonal close-packed phase of the high-entropy alloy CrMnFeCoNi

Cameron L. Tracy[1], Sulgiye Park[1], Dylan R. Rittman[1], Steven J. Zinkle[2,3], Hongbin Bei[4], Maik Lang[3], Rodney C. Ewing[1] & Wendy L. Mao[1,5]

High-entropy alloys, near-equiatomic solid solutions of five or more elements, represent a new strategy for the design of materials with properties superior to those of conventional alloys. However, their phase space remains constrained, with transition metal high-entropy alloys exhibiting only face- or body-centered cubic structures. Here, we report the high-pressure synthesis of a hexagonal close-packed phase of the prototypical high-entropy alloy CrMnFeCoNi. This martensitic transformation begins at 14 GPa and is attributed to suppression of the local magnetic moments, destabilizing the initial fcc structure. Similar to fcc-to-hcp transformations in Al and the noble gases, the transformation is sluggish, occurring over a range of >40 GPa. However, the behaviour of CrMnFeCoNi is unique in that the hcp phase is retained following decompression to ambient pressure, yielding metastable fcc-hcp mixtures. This demonstrates a means of tuning the structures and properties of high-entropy alloys in a manner not achievable by conventional processing techniques.

[1] Department of Geological Sciences, Stanford University, Stanford, California 94305, USA. [2] Department of Materials Science and Engineering, University of Tennessee, Knoxville, Tennessee 37996, USA. [3] Department of Nuclear Engineering, University of Tennessee, Knoxville, Tennessee 37996, USA. [4] Materials Science & Technology Division, Oak Ridge National Laboratory, Oak Ridge, Tennessee 37831, USA. [5] Stanford Institute for Materials & Energy Sciences, SLAC National Accelerator Laboratory, Menlo Park, California 94025, USA. Correspondence and requests for materials should be addressed to C.L.T. (email: cltracy@stanford.edu).

Conventional alloys typically consist of one or two principal component elements, to which small quantities of various alloying elements are added. Recently, a new class of materials, high-entropy alloys, was discovered[1,2]. These alloys are commonly defined as near-equiatomic solid solutions of five or more component elements[3–6]. Combining simple structures with very high-chemical disorder, high-entropy alloy design represents a new strategy for the development of materials with combinations of properties that cannot be achieved in conventional alloys. These include simultaneous ductility and strength, with the latter sometimes exceeding those of bulk metallic glasses[7,8], as well as high hardness, wear resistance, corrosion resistance and thermal stability[3–6]. Complex magnetic behaviour[9] and superconductivity[10] have also been reported. In addition to their potential use as structural materials, these alloys have been investigated for diverse applications as thermoelectric[11], soft magnetic[12] and radiation tolerant materials[13].

While high-entropy alloys encompass a wide range of compositions, their phase space remains constrained. The majority of these alloys reported to date adopt one of two simple structures: face-centered cubic (fcc) or body-centered cubic (bcc). Despite their prevalence among the transition metals, of which most high-entropy alloys are composed, hexagonal close-packed (hcp) phases are rare. High-entropy alloys with this structure, desirable due to its generally high hardness relative to more ductile fcc phases, have been reported only in the last few years. Most hcp high-entropy alloys are solid solutions of rare earth elements (Sc, Y and lanthanides), which often adopt hcp structures as elemental metals and possess relatively uniform atomic sizes and similar electronic structures[9,14–16]. Unlike transition metal high-entropy alloys, these materials lack the solid solution strengthening effect that arises from size and electronic structure mismatches among component elements[14]. Recently, the rare earth-bearing alloy LiMg$_{0.5}$AlScTi$_{1.5}$ was synthesized in an hcp phase[17], however, attempts to produce hcp high-entropy alloys without rare earth elements have been unsuccessful[18]. Similarly, recent attempts to fabricate single-phase, hcp, quaternary transition metal medium-entropy alloys resulted in compositional segregation into multiple phases[19].

Among the many high-entropy alloys reported in the past decade, the transition metal alloy CrMnFeCoNi has been extensively studied. One of the first high-entropy alloys observed[1], this fcc material exhibits ductility, high strength and toughness superior to virtually all metals and alloys[20], properties which it retains at high temperatures due to slow diffusion of its component elements[21]. This unique combination of properties results from the alloy's complex nanoscale deformation mechanisms, which include dislocation slip with strong dislocation interactions, nano-twinning and crack-tip bridging[20,22]. However, its elastic and phase behaviour during compression at high pressure has not been previously investigated.

Here we report a study of the phase space of this prototypical fcc transition metal high-entropy alloy up to a pressure of 54.1 GPa. A sluggish, martensitic transformation to an hcp phase was observed, starting at ~14 GPa. Recovery of the initial fcc phase on pressure reduction was limited, such that the new hcp high-pressure phase persists to ambient conditions. The phase fractions in the resulting fcc-hcp mixture vary with the maximum pressure achieved, such that the ratio of the two, and thus the material's properties, can be effectively tailored. These results show that high-pressure processing can make accessible phases not achievable through conventional processing methods, such as hcp transition metal high-entropy alloys. In addition, because the crystal structures and elastic properties of 3d transition metals are typically understood in terms of their d-electron occupancies and magnetic states[23], these results clarify the processes controlling phase behaviour in systems with highly mixed d-band configurations.

## Results

**High pressure phase transformation.** At ambient conditions, X-ray diffraction patterns from this alloy (Fig. 1) correspond to a single fcc phase with a unit cell parameter of $a = 3.597(1)$ Å, in agreement with the literature[1]. The unit cell contracts with increasing pressure, as indicated by shifts of the diffraction maxima to higher $2\theta$ values. Fitting of a third-order Birch–Murnaghan equation of state[24] to the refined average atomic volumes (Fig. 2) yields a bulk modulus of $B_0 = 150(3)$ GPa with a pressure derivative $B_0' = 5.1(4)$. These values are similar to those predicted by density functional theory calculations ($B_0 = 160$ GPa and $B_0' = 4.9$)[25], ab initio calculations ($B_0 = 140$ GPa)[26] and resonant ultrasound spectroscopy ($B_0 = 143$ GPa)[27], as well as to the bulk modulus of this alloy's sole fcc-structured component metal, Ni ($B_0 = 161(11)$ GPa)[28].

Starting at 14 GPa, new diffraction maxima appear. This signal is first evident in Fig. 1, as a weak peak at ~13.5° and a shoulder on the high $2\theta$ side of the fcc (111) peak. With increasing pressure these new peaks grow in intensity, while those corresponding to the initial fcc phase are attenuated. However, a small contribution from the fcc phase to the patterns remains at the highest pressures achieved. The appearance and growth of these peaks indicate a sluggish transformation of the alloy to a new phase. Refinement of the X-ray diffraction data with an hcp structure of CrMnFeCoNi yields a good fit, accounting for all pressure-induced changes to the patterns. No evidence of metallic glass or intermetallic formation was observed, indicating that the quinary solid solution remains stable, although its initial fcc structure does not.

Refinement of the unit cell parameters of the high pressure hcp phase (Fig. 2) reveals that the volume collapse accompanying the transformation is quite small, at 0.6(4)%. This is consistent with ab initio calculations predicting a 0.9% difference in average atomic volume between fcc and hcp phases of this alloy at 0 K (ref. 25). Furthermore, little change in the material's bulk modulus is evident, with fitting of a third-order Birch–Murnaghan equation of state[24] to the hcp phase atomic volumes yielding a bulk modulus of $B_0 = 141(8)$ GPa with a pressure derivative $B_0' = 5.6(2)$. Compared with the fcc structure, the hcp structure has an additional degree of freedom that can be refined, its unit cell parameter axial ratio $c/a$. The structure of the hcp phase is fully constrained by its cell parameters $c$ and $a$, the ratio of which is 1.633 in an ideal, close-packed, hard sphere system. Refinement shows that this ratio is below the ideal value when the hcp phase first forms, and that it decreases with pressure (Fig. 2, inset), indicating preferential contraction along the [001] direction.

**Transformation mechanism.** This high pressure fcc-to-hcp transformation of the high-entropy alloy CrMnFeCoNi involves two close-packed structures that differ only in the stacking sequence of their close-packed planes. Thus, the transformation requires only small atomic displacements associated with the formation of stacking faults, allowing for the local formation of the hcp phase within an fcc matrix. Because this alloy is known to have a very low stacking fault energy[26,29], the energy difference between the fcc and hcp phases is small, promoting the formation of stacking disorder during deformation. The appearance of the hcp (010) and (002) peaks before the appearance of the (011) peak, which is of greater intensity in a fully crystallized hcp phase, is consistent with a martensitic transformation mechanism based on the development of stacking disorder in the [111] direction of the initial fcc phase[30].

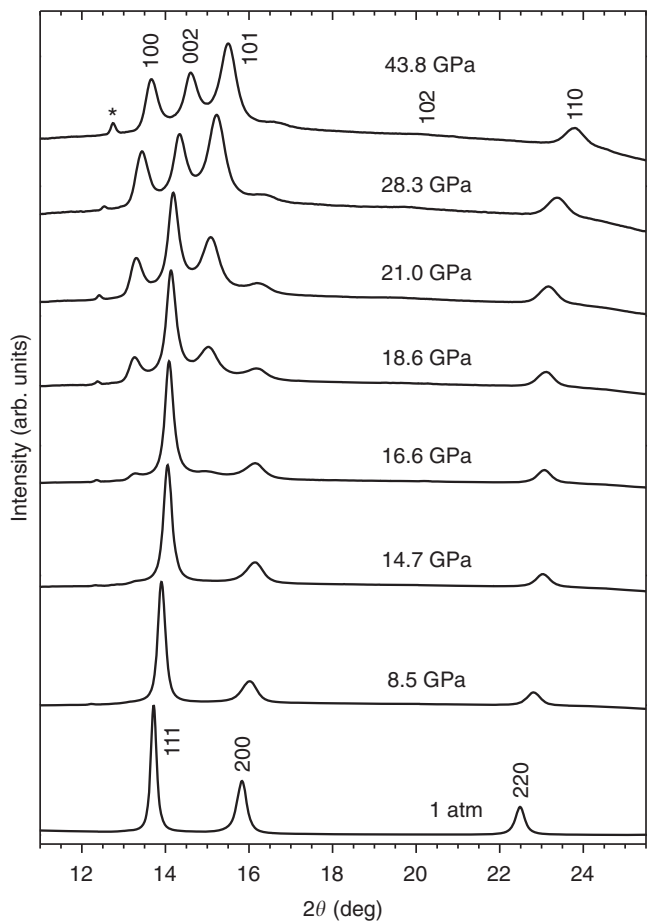

**Figure 1 | X-ray diffraction patterns of CrMnFeCoNi at various pressures.** The initial, ambient pressure pattern corresponds to an fcc structure. Starting at 14 GPa, new peaks appear and continue to grow with increasing pressure, indicating a transformation to an hcp structure. Shifts of all peaks to higher $2\theta$ with increasing pressure result from contraction of the unit cells. The peak marked with an asterisk corresponds to the Au pressure calibrant.

At ambient temperature, pressure-induced fcc-to-hcp transformations have not been previously observed in any transition metal. Those transition metal elements that exhibit an fcc structure at ambient conditions all retain this structure to the highest pressures that have been experimentally achieved[31], although an fcc-to-hcp transition has been observed in elemental Fe subjected to simultaneous high temperature and pressure[32,33]. Among all elemental materials, only the post-transition metals Al (ref. 34) and Pb (ref. 35), and the noble gases Ar (ref. 36), Kr (refs 37,38) and Xe (refs 30,38,39), exhibit ambient temperature fcc-to-hcp transformations. As with the alloys studied here, all such transformation are associated with vanishingly small volume changes, as they involve only martensitic modification of the atomic stacking order. With the exception of Pb, the fcc-to-hcp transitions in these materials are sluggish, taking place over a ~50 GPa range for Al (ref. 34) and Kr (ref. 37), a ~70 GPa range for Xe (ref. 30), and possibly a several hundred GPa range for Ar (ref. 36). This behaviour is consistent with the transformation in CrMnFeCoNi, which, as shown in Fig. 3, begins at 14 GPa and reaches 93(7)% completion at the highest pressure achieved, 54.1 GPa.

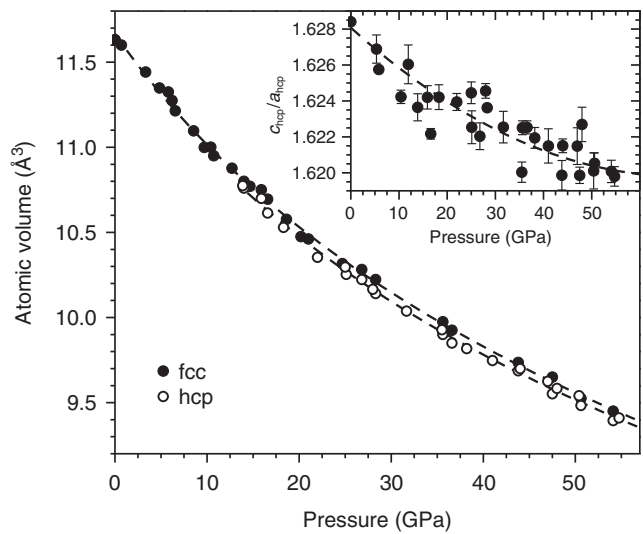

**Figure 2 | Average atomic volumes as a function of pressure.** The fcc phase is represented by closed circles, while the hcp phase is represented by open circles. The dashed lines illustrate the results of fitting of a third-order Birch–Murnaghan equation of state to the data corresponding to each phase. The inset shows the change of the unit cell axial ratio, $c/a$, with pressure. The dashed line is included to guide the eye. Error bars represent the s.d. of multiple measurements.

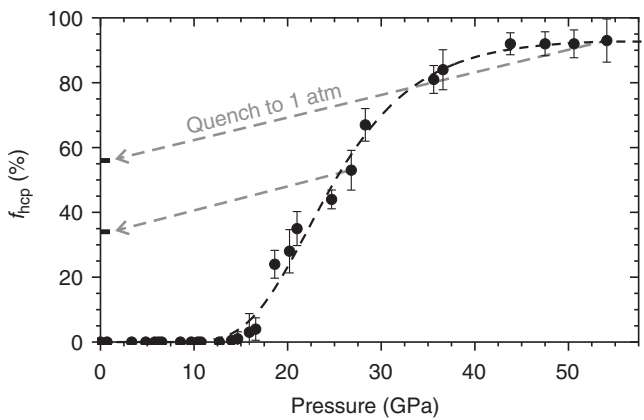

**Figure 3 | The hcp phase fraction as a function of pressure.** Starting at 14 GPa, ingrowth of the hcp phase occurs and proceeds to the highest pressure achieved, 54.1 GPa. On quenching of the sample to ambient pressure (grey arrows), most of the ingrown hcp phase is retained, with ~40% reverting to the initial fcc phase. Error bars represent the s.d. of multiple measurements.

*Ab initio* calculations have shown that the typically sluggish nature of pressure-induced fcc-to-hcp transformations results from proportionality between pressure and the energy barrier associated with stacking fault formation[40]. In this way, the pressure increase that drives the transformation also inhibits it kinetically, explaining the high pressures needed to obtain a complete conversion to the hcp phase. This suggests that the 'tail' in the phase evolution at pressures above ~35 GPa, shown in Fig. 3, might result from a very high energy barrier to stacking fault formation at these pressures, slowing the transformation of the small amount of remaining fcc material.

**Magnetic moment suppression as the driving force.** The stabilization of the hcp phase of CrMnFeCoNi, relative to the initial fcc phase, can be attributed to the effects of pressure on the magnetism of this high-entropy alloy. *Ab initio* calculations have shown that the fcc phase is stabilized, relative to an hcp structure, by magnetic contributions[25]. In the absence of magnetism, an hcp structure is predicted to be the lowest energy state of the system. Magnetism in 3*d* metals is typically suppressed by the application of pressure, due to a pressure-induced collapse of the magnetic moment[41,42]. Pressure broadens the *d* bands, reducing the density of states at the Fermi level, thus suppressing the magnetic moments of 3*d* atoms and driving the system towards a nonmagnetic state. *Ab initio* calculations confirm inverse proportionality between pressure and local magnetic moments in CrMnFeCoNi (refs 25,43). This pressure-induced suppression of the local magnetic moments in the initial fcc phase lessens the stabilizing effects of magnetism, such that formation of the hcp phase becomes energetically favourable by 14 GPa.

Similar effects of magnetic moment suppression on phase stability have been observed in the high pressure behaviour of two of this high-entropy alloy's component transition metals: Fe (refs 44–46), which undergoes fcc-to-hcp or bcc-to-hcp transformations, and Co (refs 47–53), which undergoes an hcp-to-fcc transformation. In both elemental metals, suppression of the magnetic moment lessens the stabilizing effect of magnetism on the low pressure phase, driving a transformation to a high pressure phase that is more stable in the absence of magnetic contributions. In hcp Co, the magnetic moment suppression has been identified as the cause of a decrease in the unit cell axial ratio *c*/*a* with increasing pressure[49], consistent with observation of the same unit cell distortion in the hcp phase of CrMnFeCoNi (Fig. 2, inset). The transformation of elemental Fe to the hcp phase occurs at 13 GPa (ref. 44) (at ~6% volume compression, using the bulk modulus of Guinan and Beshers[54]), in remarkable agreement with the critical pressure observed for the formation of hcp CrMnFeCoNi, 14 GPa (~7% compression).

According to *ab initio* calculations by Ma *et al.*[25], the ~1 $\mu_B$ local magnetic moment of Fe dominates the magnetism of this alloy at ambient conditions, while the other component elements have very small moments. This may be the cause of the similar critical pressures and compressions for hcp formation in Fe and CrMnFeCoNi, as transformations in both materials result primarily from the suppression of the Fe magnetic moment. In contrast to the findings of Ma *et al.*[25], similar calculations by Tian *et al.*[43] predict a local magnetic moment of Mn that is close to that of Fe at ambient conditions. However, they predict a strong sensitivity of this moment to density, such that it becomes negligible at ~4% volume compression, indicating that the magnetic moment of Fe alone is sufficient to stabilize the fcc phase of CrMnFeCoNi up to ~7% compression. This behaviour suggests that removal of Fe from this high-entropy alloy might enhance the stability of the hcp phase and reduce the critical pressure for its formation, facilitating phase manipulation.

**Retention of the hcp phase to ambient pressure.** The high pressure phase behaviour of CrMnFeCoNi differs from that of other materials that undergo fcc-to-hcp transformations in that its hcp phase is retained upon the return to ambient pressure. During depressurization, a portion of the hcp material reverts to the fcc structure, while the remainder does not undergo further transformation, persisting with ambient pressure unit cell parameters of $a = 2.544(1)$ Å and $c = 4.142(3)$ Å. This yields mixed-phase assemblages at ambient conditions (Fig. 4). As shown in Fig. 3, depressurization from 54.1 GPa, where $f_{hcp} = 93(7)\%$, yields $f_{hcp} = 56(2)\%$ at ambient conditions, while depressurization from 26.8 GPa, where $f_{hcp} = 53(6)\%$, yields $f_{hcp} = 34(1)\%$ at ambient conditions. The slope of the change in $f_{hcp}$ as a function of pressure, during pressure reduction, is remarkably consistent between the different maximum pressures achieved, indicating that it is nearly invariant as a function of $f_{hcp}$ before depressurization. In all cases, roughly 60% of the hcp volume is retained, with the remainder reverting to the fcc phase.

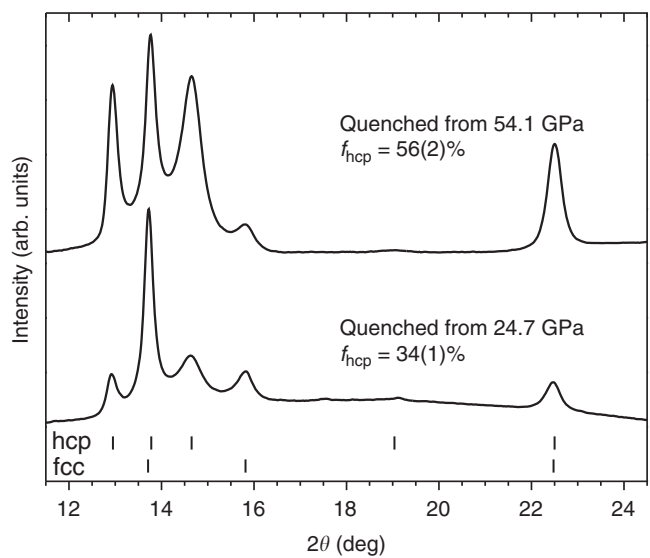

**Figure 4 | X-ray diffraction patterns of CrMnFeCoNi following decompression to ambient conditions.** Both samples exhibit a mixture of fcc and hcp phases, with $f_{hcp}$ showing proportionality with the maximum pressure achieved before quenching.

The partial recovery of the initial fcc phase with the reduction of pressure has interesting implications for its stability, relative to the hcp phase. *Ab initio* calculations[25] predict the hcp phase to be stable below 340 K, suggesting that the fcc phase is metastable at ambient temperature. This contrasts with the experimental results here, as recovery of the fcc phase during decompression at ambient temperature indicates the existence of a driving force for the hcp-to-fcc transformation, and suggests that the hcp phase may be metastable at this temperature. The partial retention of the hcp phase at ambient conditions may be related to the sluggishness of atomic motion that is common to most high-entropy alloys. This property results from the large fluctuations in lattice potential energy between sites, due to varying coordination by dissimilar component elements, which causes trapping of atoms in low energy sites[3,21]. Lattice strain resulting from the size mismatch between constituent elements further hinders the collective atomic motion needed for recovery of the fcc phase. These effects could kinetically hinder the hcp-to-fcc transformation during decompression to ambient conditions. Rapid pressure quenching might enhance the retention of the hcp phase.

The partial irreversibility of the pressure-induced fcc-to-hcp transformation, along with its sluggish progression as a function of pressure, allows for tailoring of the phase content of the fcc-hcp mixture produced following the reduction of pressure to ambient conditions, as shown in Fig. 4. Thus, high pressure processing of this material cannot only produce an hcp phase, which is rare among high-entropy alloys, but can also produce tailored mixtures of the hard hcp and ductile fcc phases without changes in composition. Such multiphase materials are attractive because they can provide uncommon combinations of high ductility, strength and toughness[55,56]. High pressure processing of this class of high-entropy alloys might allow for improvement and precise control of their critical properties.

## Methods

**Sample preparation.** An equimolar mixture of the constituent metals Cr, Mn, Fe, Co and Ni was arc-melted under an Ar atmosphere, and the resulting melt was drop cast to produce an ingot of the high-entropy alloy CrMnFeCoNi. Annealing at 1,200 °C for 24 h was performed in an evacuated quartz capsule to ensure sample homogeneity and crystallinity. The sample synthesis is described in detail elsewhere[57]. A powder was obtained by scraping the ingot with a diamond-tipped pen. The CrMnFeCoNi powder was manually loaded into a symmetric diamond anvil cell with 300 μm diameter diamond culets. The sample was contained in a W gasket along with silicone oil, as a pressure transmitting medium. Au powder was also loaded into the gasket, as a standard for determination of the cell pressure.

**X-ray diffraction.** Angle-dispersive X-ray diffraction measurements were performed at pressures of up to 54.1 GPa at beamline 16-BMD of the Advanced Photon Source[58]. An X-ray wavelength of $\lambda = 0.4959$ Å was used for all measurements. Patterns were collected on a Mar345 two-dimensional detector. Between measurements, the pressure on the diamond anvil cell was increased (or decreased, during depressurization) and the resulting pressure change was monitored by measurement of shifts in the X-ray diffraction peak positions of signal arising from the Au pressure standard. All pressure changes were performed slowly, in a stepwise manner. The sample pressure was increased no more than a few GPa between each measurement, and depressurization was performed at a rate of $\sim 1$ GPa min$^{-1}$, to avoid excessive kinetic constraints on phase formation.

**Data analysis.** Multiple X-ray diffraction measurements were performed at each pressure step. The resulting two-dimensional diffraction patterns were integrated in the azimuthal direction using the software Dioptas[59]. Rietveld refinement was performed using the software GSAS-II (ref. 60) to determine unit cell parameters and phase fractions of both the fcc and hcp phases as functions of pressure. Bulk moduli were determined from fitting of a third-order Birch–Murnaghan equation[24] to the refined atomic volumes as a function of pressure.

**Data availability.** The data that support the findings of this study are available within the article or from the corresponding author on request.

### Acknowledgements
This work was supported by the Energy Frontier Research Center 'Materials Science of Actinides' funded by the US Department of Energy, Office of Science, Office of Basic Energy Sciences (Grant No. DE-SC0001089) (high pressure experiments, C.L.T., S.P., D.R.R., M.L., R.C.E. and W.L.M.), the US Department of Energy, Office of Science, Office of Basic Energy Sciences, Materials Sciences and Engineering Division (material synthesis, H.B.), and the US Department of Energy, Office of Science, Office of Fusion Energy Sciences (Grant No. DE-SC0006661) (material synthesis, S.J.Z.). Portions of this work were performed at HPCAT (Sector 16), Advanced Photon Source (APS), Argonne National Laboratory. HPCAT operations are supported by DOE-NNSA under Award No. DE-NA0001974 and DOE-BES under Award No. DE-FG02-99ER45775, with partial instrumentation funding by NSF. The Advanced Photon Source is a US Department of Energy (DOE) Office of Science User Facility operated for the DOE Office of Science by Argonne National Laboratory under Contract No. DE-AC02-06CH11357. HPCAT beamtime was supported and allocated by the Carnegie/DOE Alliance Center (CDAC) under DOE-BES Award No. DE-NA0002006.

### Author contributions
C.L.T., M.L. and W.L.M. conceived and designed the experiments. S.J.Z. and H.B. synthesized the alloy samples. C.L.T., S.P. and D.R.R. performed the synchrotron X-ray measurements, and C.L.T. analysed the resulting data. The manuscript was drafted by C.L.T., R.C.E. and W.L.M. All authors discussed and commented on the manuscript.

### Additional information
**Competing interests:** The authors declare no competing financial interests.

**Reprints and permission** information is available online at http://npg.nature.com/reprintsandpermissions/

**How to cite this article:** Tracy, C. L. *et al.* High pressure synthesis of a hexagonal close-packed phase of the high-entropy alloy CrMnFeCoNi. *Nat. Commun.* **8**, 15634 doi: 10.1038/ncomms15634 (2017).

**Publisher's note**: Springer Nature remains neutral with regard to jurisdictional claims in published maps and institutional affiliations.